\documentclass[iop]{emulateapj}

\slugcomment{ApJ, in press}

\shorttitle{Secondary Origin for Cosmic Ray Positrons}
\shortauthors{Kruskal, Ahlen and Tarl\'e}
\usepackage{natbib}
\usepackage{IEEEtrantools}
\usepackage{amsmath}
\usepackage{array}
\usepackage{csquotes}
\usepackage{graphicx}
\usepackage[percent]{overpic}
\usepackage{bm}

\begin{document}
\bibliographystyle{apj}

\title{Secondary Production as the Origin for the Cosmic Ray Positron Excess}

\author{M. Kruskal, S.P. Ahlen}
\affil{Physics Department, Boston University, Boston, Massachusetts 02215, USA}
\and
\author{G. Tarl\'e}
\affil{Physics Department, University of Michigan Ann Arbor, MI 48109, USA}

\begin{abstract}
The Alpha Magnetic Spectrometer has released high-precision data for cosmic rays, and has verified an excess of positrons relative to expectations from cosmic ray interactions in the interstellar medium. An exciting and well-known possibility for the excess is production of electron-positron pairs by annihilating dark matter particles in the halo of the Galaxy. We have constructed a new scenario for propagation of cosmic rays, based on the 2000 SMILI results and various other astrophysics observations and measurements, in which the positron excess is due to secondary production. The scenario is studied from a simple heuristic perspective, and also within the constraints of a diffusion-reacceleration model using GALPROP. The conclusions of each approach agree with one another, showing that the scenario agrees well with the observed positron flux, without any need for dark matter or other exotic production mechanisms.   
\end{abstract}


\section{INTRODUCTION}

Ever since the seminal search in 1981 for low energy antiprotons in the cosmic rays \citep{Buff} and the interpretation of these results in terms of annihilating dark matter \citep{Silk} cosmic ray antimatter has figured prominently in searches for dark matter. Although antiprotons have been shown to be due to secondary production, positrons have been more interesting. The HEAT Collaboration published results from 1995 \citep{HEAT95} through 2004 \citep{HEAT04} on the positron fraction $e^+/(e^+ + e^-)$ that indicated an excess from 10--50 GeV when compared to a standard propagation model \citep{Strong}.  PAMELA, with a larger exposure than HEAT, confirmed the excess and showed continued growth to 100 GeV \citep{PAMELA09}.  PAMELA did not find the alleged ``smoking gun" of dark matter---an abrupt decrease of positrons at high energy. With greater than three years on the International Space Station, AMS has collected far more events than HEAT or PAMELA. The AMS-02 positron fraction \citep{AMS14Accardo} appears to level off or possibly peak near 275 GeV. The AMS-02 positron differential flux F \citep{AMS14} follows a power law spectrum from 40--430 GeV, F = F$_0$E$^{-\alpha}$ with $\alpha$= 2.78. This value is very close to the spectral index of protons up to 1 TeV, suggesting a connection between positrons and primary protons. The spectral index for electrons is much larger (between 3.1 and 3.3), which may reflect features related to the acceleration and escape of primary cosmic ray electrons from their sources, and/or to energy loss by synchrotron radiation and inverse Compton scattering in the sources and interstellar medium (ISM).

There are a number of possible explanations for the positron excess, the most exciting of which is the annihilation of dark matter particles in the halo of the Galaxy. The absence of an antiproton excess would then suggest that dark matter must couple more strongly to leptons than to quarks or gauge bosons \citep{Cirelli}.  The required annihilation cross section is quite large, and requires Sommerfeld-type enhancements to increase the annihilation rate at halo velocities which are much smaller than velocities at freeze-out. A comparison of a recent leptophilic model with AMS data \citep{Cao} yields a best-fit dark matter particle mass of 600 GeV with best-fit values of $\left<\sigma v\right> = 9 \times 10^{-24} \:\textrm{cm}^3\:\textrm{s}^{-1}$ and branching ratios of 65\%, 17.5\%, and 17.5\% for $\tau^+\tau^-$, $\mu^+\mu^-$, $e^+e^-$ final states respectively. This cross section is impossible to reconcile with Fermi-LAT observations of dwarf satellite galaxies \citep{Ackermann}, which have placed a limit of $\left<\sigma v\right> < 10^{-24} \:\textrm{cm}^3\:\textrm{s}^{-1}$ at a dark matter mass of 600 GeV for the $\tau$ final state. Observations of the Cosmic Microwave Background (CMB) also constrain annihilation cross sections. The WMAP limit \citep{Slatyer} is $\left<\sigma v\right> < 2 \times 10^{-24} \:\textrm{cm}^3\:\textrm{s}^{-1}$  for a dark matter mass of 600 GeV. Another popular explanation for the excess is positron production in pulsar wind nebulae (PWN) \citep{DiMauro}. Large numbers of $e^\pm$ pairs can be produced through a complex sequence of events originating with the extraction of an electron from the surface of a neutron star. It is believed these particles are released into the ISM in a burst after a time of order 50 kyr. The details by which the electrons and positrons are released are not well known, so there are large uncertainties regarding their flux and spectra. There are many free parameters, and it is likely that almost any observed positron spectrum could be fit to a PWN model.

The essence of any experimental search for new physics is the discrimination of signal from background. In the case of the search for dark matter annihilation or PWN particle acceleration through excesses in the positron flux, the background is secondary production, namely production of positrons in collisions of cosmic ray nuclei with nuclei of the ISM. The most common such collision is between cosmic ray protons or helium nuclei with hydrogen or helium nuclei. A great deal is known about such collisions from measurements of cross sections and branching ratios at particle colliders and accelerators. The possibility that all cosmic ray positrons are produced in cosmic ray collisions has been considered by various authors. \citet{BlumPRL} proposed that the AMS-02 positron fraction was consistent with a secondary origin, provided the galactic confinement time $\tau \geq 30$ Myr at 10 GeV, and $\tau \leq 1$ Myr at 300 GeV. They also found that the mean atomic number density $n \geq 1$ cm$^{-3}$ at 300 GeV. Cowsik and collaborators \citep{Cowsik10,Cowsik14} used the Nested Leaky Box model to conclude that the secondary production of all positrons was possible if $\tau \approx 2$ Myr, independent of energy, and if $n = 0.5 \textrm{ cm}^{-3}$. In this paper we will show that a secondary origin of positrons can be obtained by considering observations involving other low-mass secondary particles in the cosmic rays:

\begin{enumerate}
\item	The ratio of radioactive  $^{10}$Be to stable  $^{9}$Be will be used to constrain the confinement time of light secondary cosmic rays. This will allow us to show that energy loss of secondary positrons due to synchrotron radiation and inverse Compton scattering in the ISM is small for energies below a few hundred GeV.
\item	The flux of antiprotons will be used to determine the path-length (in g\:cm$^{-2}$), which we shall refer to as ``grammage,'' between the primary proton source and the Earth as a function of energy. It will be found that this is consistent with the grammage determined from the B/C ratio above 1 GeV/nucleon.
\end{enumerate}

Knowledge of grammage, proton energy spectrum and properties of primary cosmic ray interactions with nuclei of the ISM will allow us to calculate the secondary positron flux and show that it agrees with measurements. We will do this in two ways, first with a simple estimate based on the positron data and source functions of secondary positron and antiproton production. We will then use the computer program GALPROP \citep{Strong} to carry out detailed diffusion-reacceleration calculations of cosmic ray transport for a set of parameters that fit the observations.

\section{BERYLLIUM and COSMIC RAY CONFINEMENT TIME}

At low energies, measurements of the abundance of the beta decay isotope $^{10}$Be can be used to measure cosmic ray confinement time $\tau$. Three isotopes of beryllium are sufficiently 
stable to be observed in the cosmic rays: $^7$Be decays by electron capture with a half-life of 53 days, but is stable in the cosmic rays since it moves at high velocity and its orbital electrons have been fully stripped; $^9$Be is stable; $^{10}$Be decays by beta emission to the stable isotope $^{10}$B with a half-life of 1.39 Myr. 
\begin{table*}[!htbp]
	\centering
	\renewcommand{\arraystretch}{\numexpr\linewidth/\columnwidth}
	\begin{tabular}{|>{\centering\arraybackslash}m{8.2em}|>{\centering\arraybackslash}m{6em}|>{\centering\arraybackslash}m{5em}|>{\centering\arraybackslash}m{10.3em}|} \hline
	REACTION & C or O BEAM ENERGY (GeV/nucleon) & CROSS SECTION (mb) & REFERENCE \\ \hline
	$^{12}$C + p $\rightarrow$ $^7$Be & 1.05 & $8.45 \pm 0.81$ & \citep{Lindstrom} \\
	$^{12}$C + p $\rightarrow$ $^9$Be & 1.05 & $5.13 \pm 0.54$ & \citep{Lindstrom} \\
	$^{12}$C + p $\rightarrow$ $^{10}$Be & 1.05 & $3.41 \pm 0.35$ & \citep{Lindstrom} \\
	$^{12}$C + p $\rightarrow$ $^7$Be & 2.1 & $9.49 \pm 0.99$ & \citep{Lindstrom} \\
	$^{12}$C + p $\rightarrow$ $^9$Be & 2.1 & $5.92 \pm 0.54$ & \citep{Lindstrom} \\
	$^{12}$C + p $\rightarrow$ $^{10}$Be & 2.1 & $3.42 \pm 0.35$ & \citep{Lindstrom} \\
	$^{16}$O + p $\rightarrow$ $^7$Be & 2.1 & $10.1 \pm 0.12$ & \citep{Lindstrom} \\
	$^{16}$O + p $\rightarrow$ $^9$Be & 2.1 & $4.17 \pm 0.55$ & \citep{Lindstrom} \\
	$^{16}$O + p $\rightarrow$ $^{10}$Be & 2.1 & $2.05 \pm 0.31$ & \citep{Lindstrom} \\
	$^{12}$C + p $\rightarrow$ $^7$Be & 3.66 & $10.1 \pm 1.3$ & \citep{Korejwo} \\
	$^{12}$C + p $\rightarrow$ $^9$Be & 3.66 & $6.7 \pm 0.9$ & \citep{Korejwo} \\
	$^{12}$C + p $\rightarrow$ $^{10}$Be & 3.66 & $4.2 \pm 0.6$ & \citep{Korejwo} \\ \hline
	\end{tabular}
	\caption{Cross sections for production of beryllium isotopes.}
	\label{tab:xsecBe}
\end{table*}
Beryllium is produced predominantly by collisions of primary carbon and oxygen cosmic ray nuclei with nuclei of the ISM. Beryllium isotope production cross sections for carbon and oxygen collisions with protons have been measured at accelerators at several energies, and are given in Table \ref{tab:xsecBe}.  Additional data can be found in \citet{Tomassetti}. If there is no decay of $^{10}$Be, the ratio $^{10}$Be/$^{9}$Be $\approx$ 0.6.

\begin{figure}[!h]
	\centering
	\includegraphics[width=0.98\columnwidth]{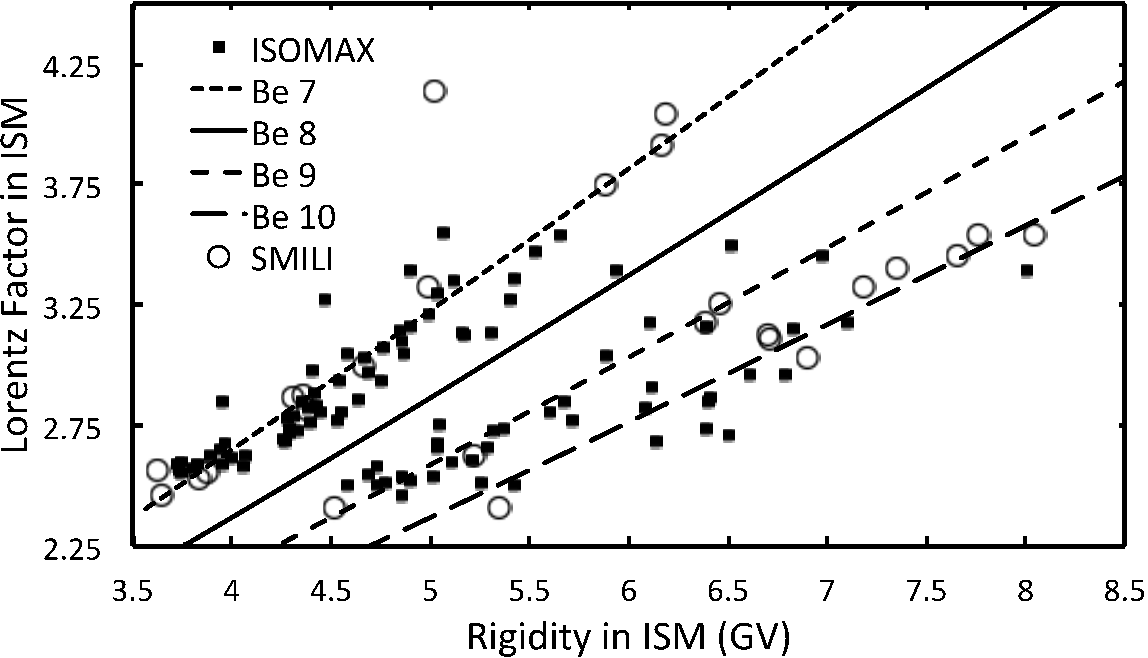}
	\caption{Scatter plot of beryllium isotope cosmic ray measurements from ISOMAX \citep{deNolfo} and SMILI \citep{SMILI}. Lorentz factor and rigidity have been corrected for solar modulation and correspond to values in the ISM of each cosmic ray.  The solar modulation potentials used were 568 MV for ISOMAX and 1938 MV for SMILI.}
	\label{fig:dataBe}
\end{figure}

CRIS \citep{Yanasak} has measured the abundance of beryllium isotopes with a satellite dE/dx telescope and has found 13 Myr $< \tau <$ 16 Myr for energies near 100 MeV/nucleon. This energy is quite low, and the results may not be representative of the confinement times of higher rigidity cosmic rays such as for the positrons of interest here.
\begin{figure*}[!ht]
	\centering
  \includegraphics[width=0.98\textwidth]{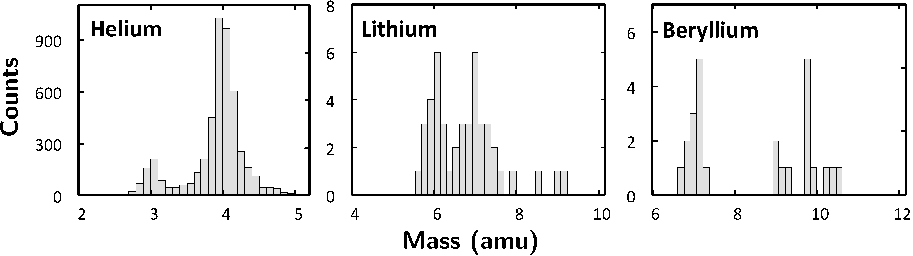}
	\caption{Mass histograms at the instrument for helium, lithium and beryllium for SMILI \citep{SMILI, Loomba}.}
	\label{fig:SMILImasses}
\end{figure*}

\begin{figure*}[!ht]
	\centering
  \includegraphics[width=0.98\textwidth]{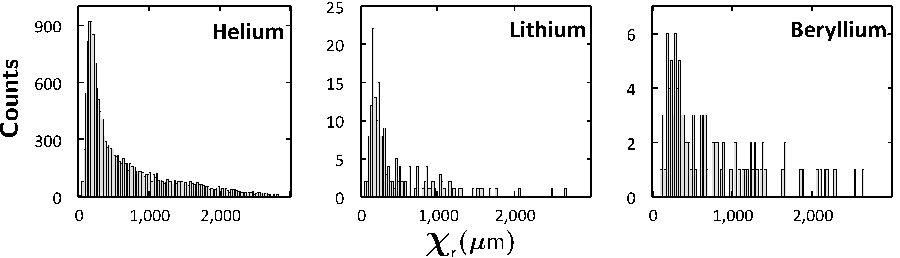}
	\caption{Track rms residual distributions for track fitting for SMILI \citep{SMILI}.}
	\label{fig:SMILIresiduals}
\end{figure*}

SMILI \citep{SMILI} and ISOMAX \citep{Hams} have measured beryllium isotope composition with balloon-borne superconducting magnetic spectrometers. SMILI was flown July 24, 1991 for 22 hours, at one of the highest levels of solar activity ever recorded, with a solar modulation constant of 1938 MV \citep{Usoskin}.  ISOMAX was flown August 4, 1998 
for 29 hours, at low solar activity with a modulation constant of 568 MV \citep{Usoskin}. A scatter plot of the Lorentz factors vs.~rigidities of detected Beryllium cosmic rays for the ISOMAX Cherenkov data \citep{deNolfo} and for the SMILI time-of-flight data are shown in Figure \ref{fig:dataBe}.  The force field approximation of \citet{Usoskin} has been used to estimate rigidity values in the ISM.

To address possible concerns regarding the mass resolution of SMILI, we show in Figure \ref{fig:SMILImasses} the mass histograms from SMILI for helium, lithium and beryllium. The events with masses $> 7$ amu for Li were most likely caused by incorrect track reconstruction due to delta rays produced in the drift tubes. Track rms residuals are shown in Figure \ref{fig:SMILIresiduals}, defined in \citet{Loomba} as,
\begin{equation}
	\chi_r^2 \equiv \dfrac{\sum^{n_x}_{i=1}\left(x_i\right)^2 + \sum^{n_y}_{i=1}\left(y_i\right)^2}{n_x+n_y-5},
\end{equation}
where $n_x$ and $n_y$ are the number of drift tubes hit in each of the two orthogonal detection planes, and $x_i$ and $y_i$ are their respective track residuals.  These indicate that delta-ray production, which is responsible the long residual tails, was similar for helium and lithium, and only slightly worse for beryllium.  Requirements were placed on the measured velocity, rms track residuals, and the consistency of signals in different scintillators in order to maximize mass resolution, reduce contamination from nuclear reactions, and reduce delta ray effects, respectively.  The resulting efficiencies for helium, lithium and beryllium were 41\%, 26\% and 27\% respectively. Observation of the lithium spectrum suggests that the spillover of $^9$Be into the $^{10}$Be mass band due to tracking errors occurred for no more than about 10\% of the $^9$Be events. Therefore, this mechanism could not have caused the large number of $^{10}$Be events observed in SMILI, relative to $^9$Be events.

By assigning mass to the nearest acceptable integer mass curve in Figure \ref{fig:dataBe}, one finds 67 $^7$Be, 28 $^9$Be and 24 $^{10}$Be events.  Some of these were made in the $\sim$5 g/cm$^2$ of residual atmosphere.  We make a rough estimate of the atmospheric effects by using cross sections at 2.1 GeV/nucleon from \citet{Lindstrom}.   We find the numbers of mean path-lengths for producing beryllium isotopes in the atmosphere by CNO cosmic rays to be 0.00471, 0.00221, and 0.00107 for $^7$Be, $^9$Be and $^{10}$Be respectively. The corresponding numbers for production in the ISM are 0.0590, 0.0304 and 0.0166.  By using the equation for total inelastic cross section from \citet{Bradt}, we find the survival probabilities for $^7$Be, $^9$Be and $^{10}$Be in the ISM to be 0.548, 0.485 and 0.458 respectively.  The corresponding survival probabilities in the atmosphere are 0.868, 0.855 and 0.850.  Assuming no significant loss of total numbers of CNO nuclei through fragmentation, one can use these numbers to estimate the number of each beryllium isotope above the atmosphere. The number of $^7$Be isotopes is 67/[0.868 + (0.00471/0.059)($\sqrt{0.868}$/0.548)] = 66.8. Similarly, the numbers of $^9$Be and $^{10}$Be isotopes above the atmosphere are 28.2 and 24.7 respectively. We see that the effects of the atmosphere are very small. This is confirmed by the more accurate calculations in \citet{SMILI}. We should note that interactions in the instrument are not of as great concern as in the atmosphere since most of these can be identified by their changes in charge or directions of motion.

The average energy in the ISM of the SMILI events is 2.0 GeV/nucleon and that of the ISOMAX events is 1.7 GeV/nucleon. Thus, including the effects of time dilation, the $^{10}$Be half-life is about 3.9 Myr. In the absence of decay, one would expect around 19 $^{10}$Be events for every 67 $^{7}$Be events, which is consistent with the data. For a mean cosmic ray transport time of 2 Myr, one would expect 13.4 $^{10}$Be events, which requires a three standard deviation to get to the observed number of 24.7. Thus, the high energy beryllium data imply that the mean confinement time for cosmic rays must be no more than 2 Myr, an order of magnitude smaller than the time determined by measurements at 100 MeV/nucleon.

\section{ENERGY LOSS of ELECTRONS and POSITRONS}

Electrons and positrons lose energy while moving through the ISM due to synchrotron radiation from interaction with the Galactic magnetic field, and due to inverse Compton scattering of cosmic microwave and starlight photons. If we ignore ionization energy loss, the electron/positron energy loss can be described by the equation \citep{Cowsik10,Cowsik14}:

\begin{equation}
	\dfrac{dE}{dt} = -bE^2
\end{equation}
where E is the positron or electron energy and b = $1.60 \times 10^{-3}$ Myr$^{-1}$GeV$^{-1}$. The initial energy $E_0$ can be expressed in terms of the energy E at time t:

\begin{equation}
	E_0=\dfrac{E}{1-btE}
\end{equation}

If the differential source spectrum is dN/d$E_0$ = K$E_0^{-\alpha}$, the spectrum at time t is:

\begin{equation}
	\dfrac{dN}{dE} = \dfrac{K}{E^\alpha}\left(1-btE\right)^{\alpha-2}
\end{equation}

The flux is zero for $E > (bt)^{-1}$. Secondary positrons at the Earth have a continuous distribution of t ranging from zero to millions of years, with a mean value of about 1--2 Myr based on the SMILI beryllium measurement. Electrons, which are dominated by a primary component, will have a different time distribution since the spatial distribution of primary electron sources (which are probably supernovae remnants) is not continuous. Since secondary positrons can be produced very shortly before being detected, and the smallest electron arrival time is determined by the distance to the nearest source, it is likely that the secondary positron energy spectrum is affected less severely than that of the primary electrons. 

The shape of the secondary positron spectrum depends on the details of production and propagation which we will treat later for different models with the use of GALPROP. For the present, we illustrate the effect on spectral index by assuming a simple distribution of survival times. Suppose 1/2 of the positrons arrive with t = 0 Myr, 1/4 arrive with t = 1 Myr, and 1/4 arrive with t = 4.33 Myr. This would give a mean time of 1.33 Myr. If we assume a source spectrum with a spectral index of 2.6, the weighted sum of the three energy spectra for the different survival times for 50--500 GeV shows the spectral index increasing from approximately 2.6 to 2.8.

\section{ANTIPROTON and POSITRON SOURCE FUNCTIONS}

Calculations of secondary antiprotons and positrons at the Earth rely on measurements and calculations of pp reaction data, and on the cosmic ray proton energy spectrum in the ISM. This spectrum follows power laws over many orders of magnitude of energy, with abrupt shifts in spectral index $\alpha_p$ at several energies. Fermi-LAT \citep{AckermannPRL} used gamma ray emission from the Earth's limb to study the primary proton spectrum. They found the proton spectral index $\alpha_p = 2.81 \pm 0.10$ before a break at 276 GeV, and $2.60 \pm 0.08$ after the break. These results were compared \citep{AckermannPRL} with direct measurements of protons from a variety of experiments (ATIC, CREAM, BESS, PAMELA, AMS-01) and found to be in fairly good agreement. The spectral break has been confirmed recently by AMS-02 \citep{AMS15}, which observes a shift of index from 2.83 at 100 GV to 2.72 at 1000 GV.

\begin{figure}[!htbp]
	\centering
	\includegraphics[width=0.98\columnwidth]{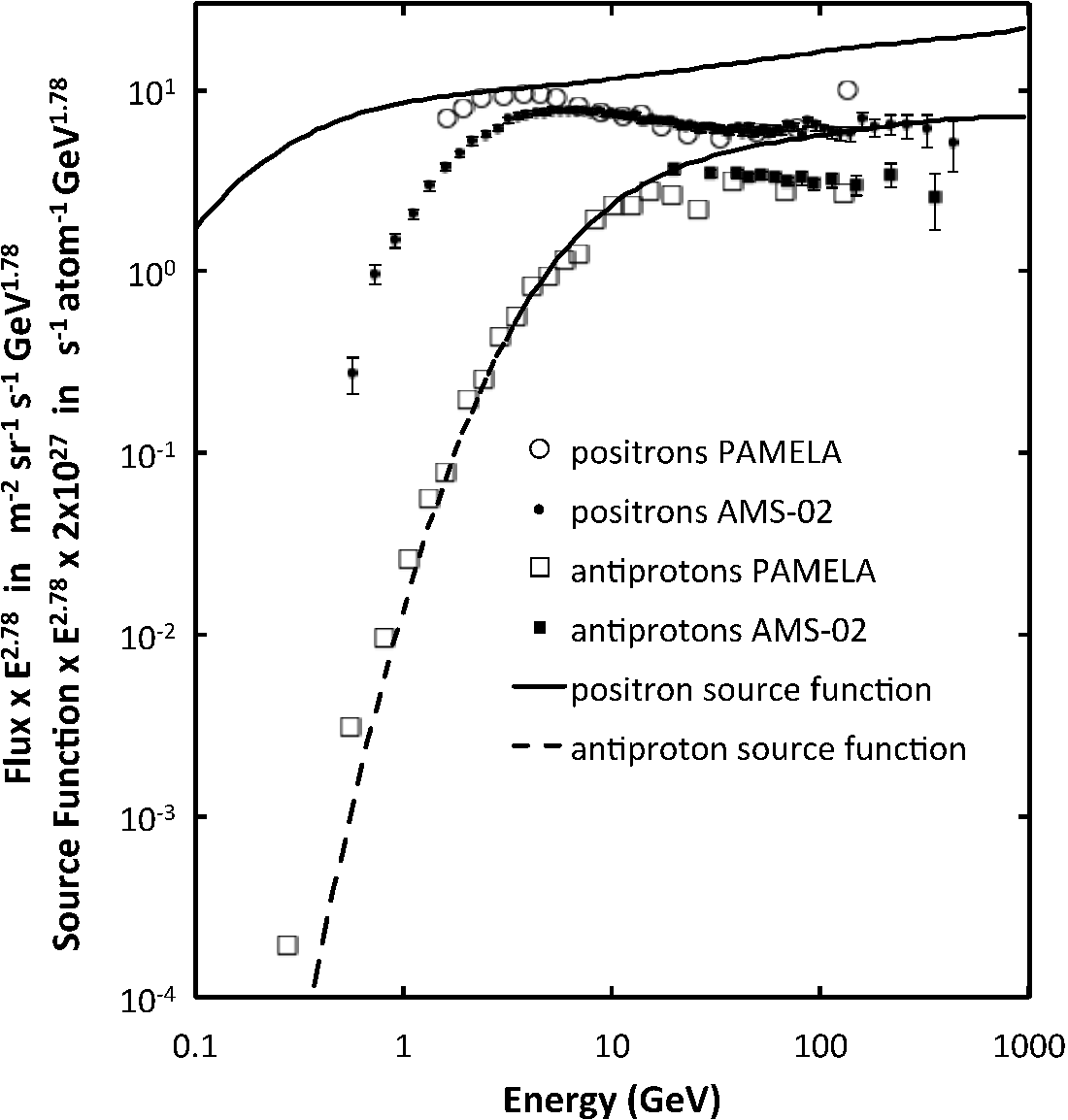}
	\caption{Antiproton and positron source functions, and flux data for cosmic ray antiprotons and positrons. The source functions include the effects of proton and helium cosmic ray particles, and hydrogen and helium atoms in the ISM.  The antiproton source function has been taken from \citet{Kappl}, and the positron source function has been taken from \citet{Protheroe}.  Both of these have been scaled for comparison to the observed flux measurements. Positron data are from \citet{PAMELA13} and \citet{AMS14}. Antiproton data are from \citet{PAMELA10} and \citet{Kounine}}.
	\label{fig:sources}
\end{figure}

The source function q is defined to be the production rate per time, per atom, per energy of the secondary particle. We show in Figure \ref{fig:sources} the antiproton and positron source functions. Each of these includes the effects of proton and helium cosmic ray particles, and hydrogen and helium atoms in the ISM. The antiproton source function has been taken from \citet{Kappl}. The positron source function has been taken from \citet{Protheroe}.  We have chosen the Protheroe source function due to its assumption of a proton spectral index of 2.6, which is appropriate for the protons responsible for production of positrons with energy above 100 GeV. We have modified Protheroe's result by using an improved nuclear enhancement factor including the effects of helium from \citet{Mori}.

We also show positron and antiproton data in Figure \ref{fig:sources}. Positron data are from PAMELA \citep{PAMELA13} and AMS-02 \citep{AMS14}. The PAMELA data are for a period of low solar activity, while the AMS-02 data are for a period of high solar activity. This accounts for the discrepancy at low energy, where flux is influenced strongly by solar modulation. Antiproton data are from PAMELA \citep{PAMELA10} and AMS-02 \citep{Kounine}.

The mean proton grammage $X$ can be estimated from Figure \ref{fig:sources} assuming the antiprotons are made by secondary production. It is easily shown that $X = 4\pi mF/q$, where F is the flux, q is the source function, and m is the mean atomic mass, which is very nearly the same as the hydrogen mass. For antiproton energy from 1--10 GeV, $X \approx$ 4 g\:cm$^{-2}$ is nearly constant. This reduces to about 2 g\:cm$^{-2}$ at 300 GeV. Note that proton parents of antiprotons have several times greater energy than the antiprotons due to the fact that initial kinetic energy is shared among many secondary particles. Taking this into account, the range of proton grammage is roughly consistent with that determined by the B/C ratio which varies from about 10 g\:cm$^{-2}$ at 1 GeV/nucleon to 2.5 g\:cm$^{-2}$ at 300 GeV/nucleon.

It is important to note that the grammage inferred from the positron data from 1--5 GeV is within 10\% of that determined by the antiproton data below 10--20 GeV. Since synchrotron and inverse Compton energy losses are not expected for such low energy positrons, this shows that low energy positrons are due to secondary production. Furthermore, if one assumes no energy loss for positrons up to a few hundred GeV, the positron data agree with the antiproton data that the grammage is reduced by a factor of two as positron energy increases up to 40 GeV. Beyond 40 GeV there is a break in the positron power law spectrum, which suggests that grammage reduces more slowly with energy, if at all, above that energy. Since a positron's proton parent typically has an order of magnitude greater energy than the positron, this shows that grammage becomes insensitive to energy for protons above 1 TeV. Beyond the break at 40 GeV, the positron spectral index is larger than 2.6 by about 0.2, suggesting the possibility of energy loss effects of the same amount as estimated in the previous section.

To conclude this section, we summarize the essential fundamental principles upon which our scenario is based:

\begin{itemize}
	\item Cosmic ray confinement time is about 1--2 Myr based on beryllium isotope measurements with rigidity from 5--8 GV. It is unlikely that the lifetime changes much with increasing rigidity, since this would make the observed anisotropy measurements incompatible with simple calculations \citep{Cowsik10,Cowsik14}.
	\item Confinement time is about 10 times longer for rigidity of 1--2 GV. As discussed below, this dramatic result cannot be understood from simple models involving diffusion and re-acceleration, and therefore is not easy to replicate with GALPROP. It is interesting to note that the deviation of the antiproton flux from the source function curve in Figure \ref{fig:sources} suggests the antiproton confinement time is reduced by a factor of 10 as rigidity increases from 1--3 GV, and remains constant above 3 GV. This effect is not related to grammage variation since such low energy antiprotons are all probably made by protons with energies slightly above the 6 GeV threshold energy.
	\item The rigidity dependent grammage distribution determined by antiproton measurements is consistent with that determined from B/C measurements, showing that antiprotons are secondary particles.
	\item The break of the positron power law spectrum at 40 GeV suggests that the diffusion coefficient becomes insensitive to rigidity above 1 TV. This could be responsible for the break in the proton spectral index above several hundred GV.
	\item The difference between the positron spectral index from 40--500 GV and that of the proton spectrum above 1 TV suggests that positron energy loss increases the positron spectral index only slightly from 2.6 to 2.8.
\end{itemize}

\section{GALPROP RESULTS}

In this section we consider a GALPROP based diffusion-reacceleration model that incorporates these features and show that it agrees with proton, beryllium, B/C, antiproton and positron data for rigidities above 3 GV.  GALPROP, version 54.1.984\footnote{Note that an updated version of GALPROP can be found at http://sourceforge.net/projects/galprop incorporating many new developments.} used here, is a program that numerically solves the transport equations for cosmic rays that are produced in the Galaxy and then propagate in the ISM. GALPROP makes use of detailed astrophysics, nuclear physics, and particle physics measurements and theory pertaining to such phenomena as Galactic magnetic fields, gas composition and density of the ISM, nuclear fragmentation cross sections, positron and antiproton source functions, etc. It is very flexible, allowing for the specification of numerous parameters such as source intensity and energy spectra for primary particles, the diffusion coefficient and its energy dependence, Alfv\'en velocity, Galactic wind velocity, etc. It allows one to treat cosmic ray propagation as purely diffusive, or with the combination of diffusion and re-acceleration.
We have considered many models within the framework of GALPROP. The challenge is to find a model that allows for a short cosmic ray confinement time to be consistent with measurements of the $^{10}$Be/$^9$Be ratio at a few GeV/nucleon, a long confinement time required for the $^{10}$Be/$^9$Be ratio at a few hundred MeV/nucleon, and the proper amount of grammage to be consistent with measurements of the B/C ratio and antiproton flux at all energies. If one can find such a model, and if that Model predicts a secondary positron flux in agreement with data, then one may conclude that observations do not require additional sources of cosmic ray positrons.

\begin{figure*}[!htbp]
	\centering
	\begin{overpic}[width=0.38\textwidth]{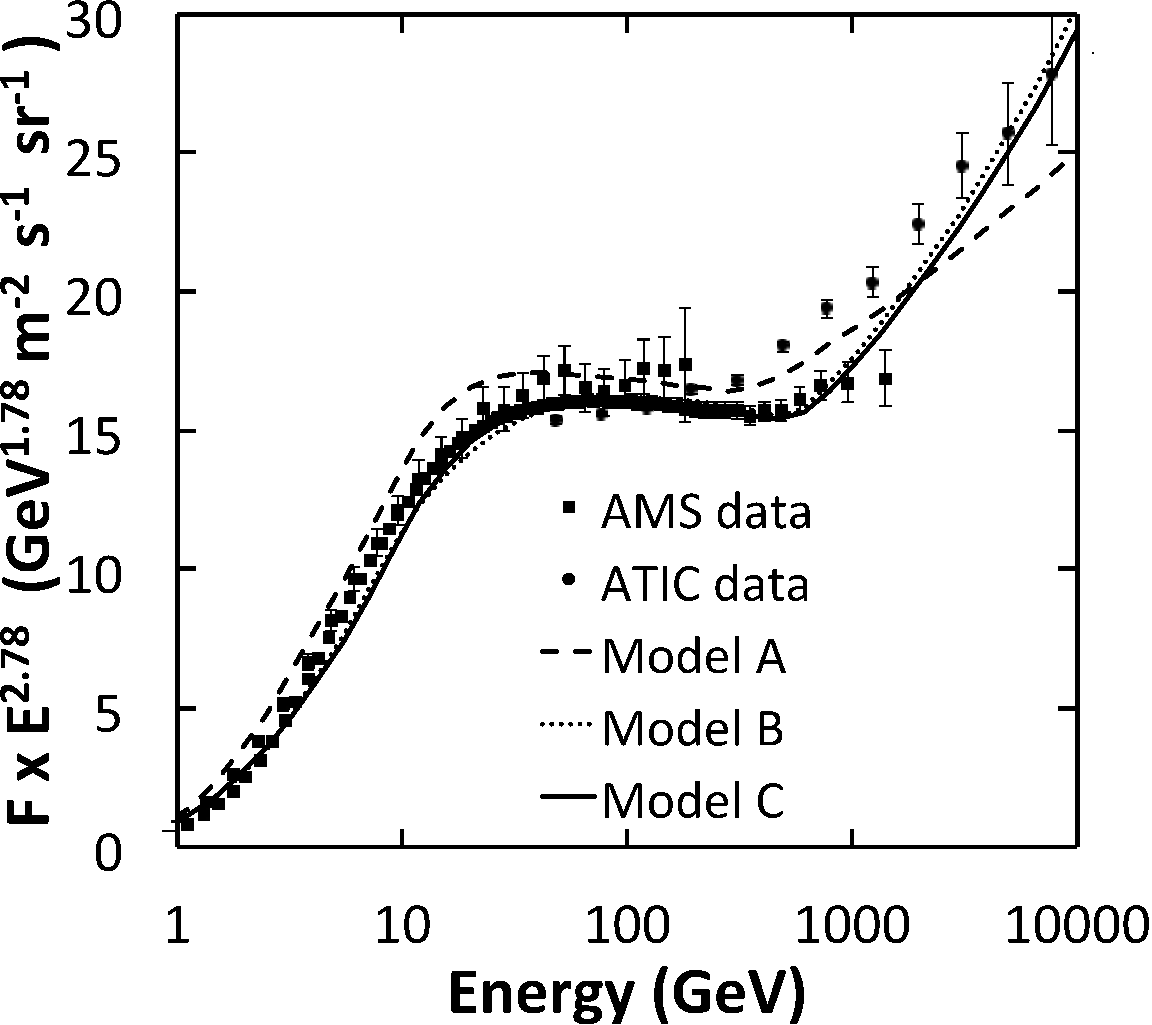}
		\put(25,75){\textbf{\textsf{Protons}}}
	\end{overpic}
	\begin{overpic}[width=0.38\textwidth]{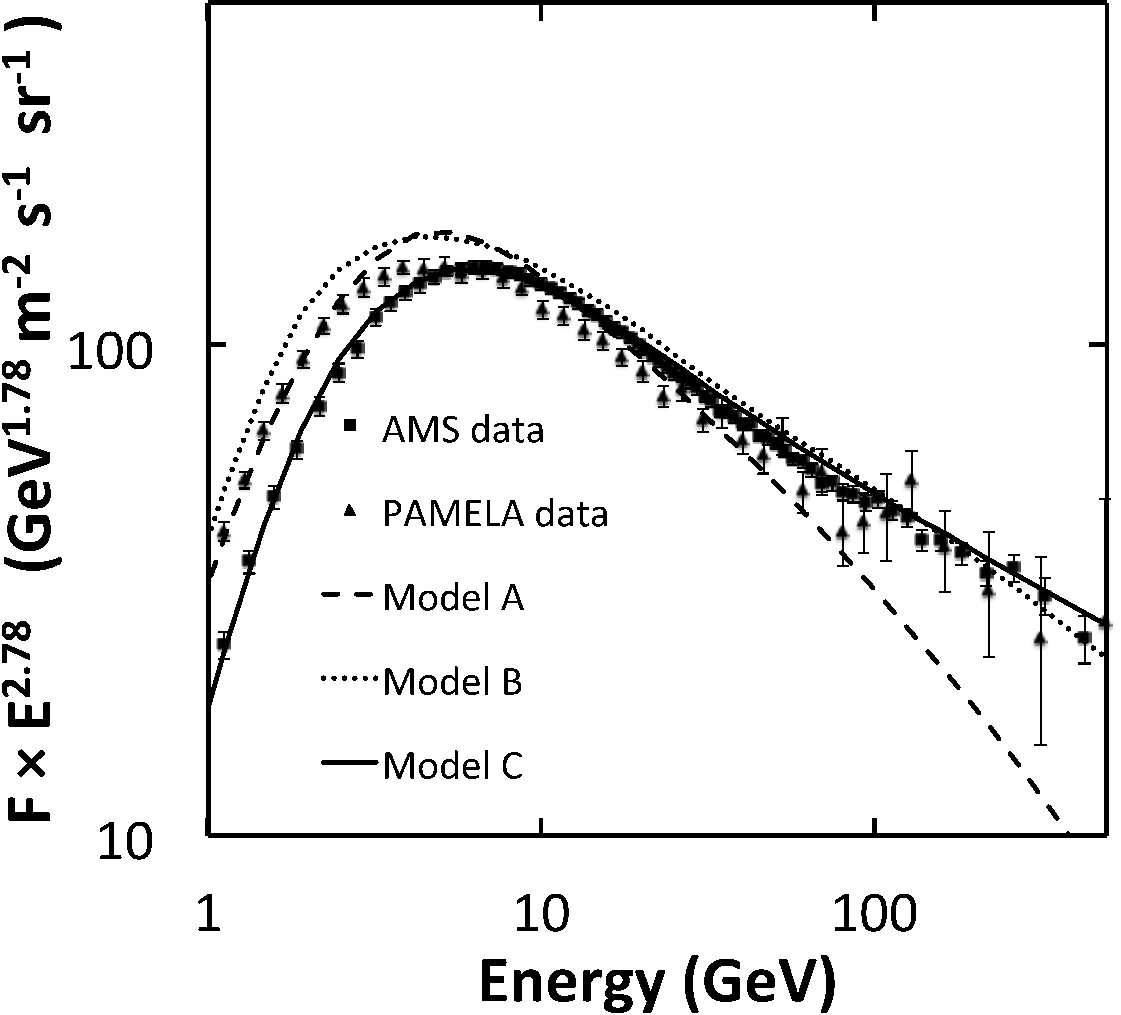}
		\put(50,75){\textbf{\textsf{Electrons}}}
	\end{overpic}
	\\
	\begin{overpic}[width=0.38\textwidth]{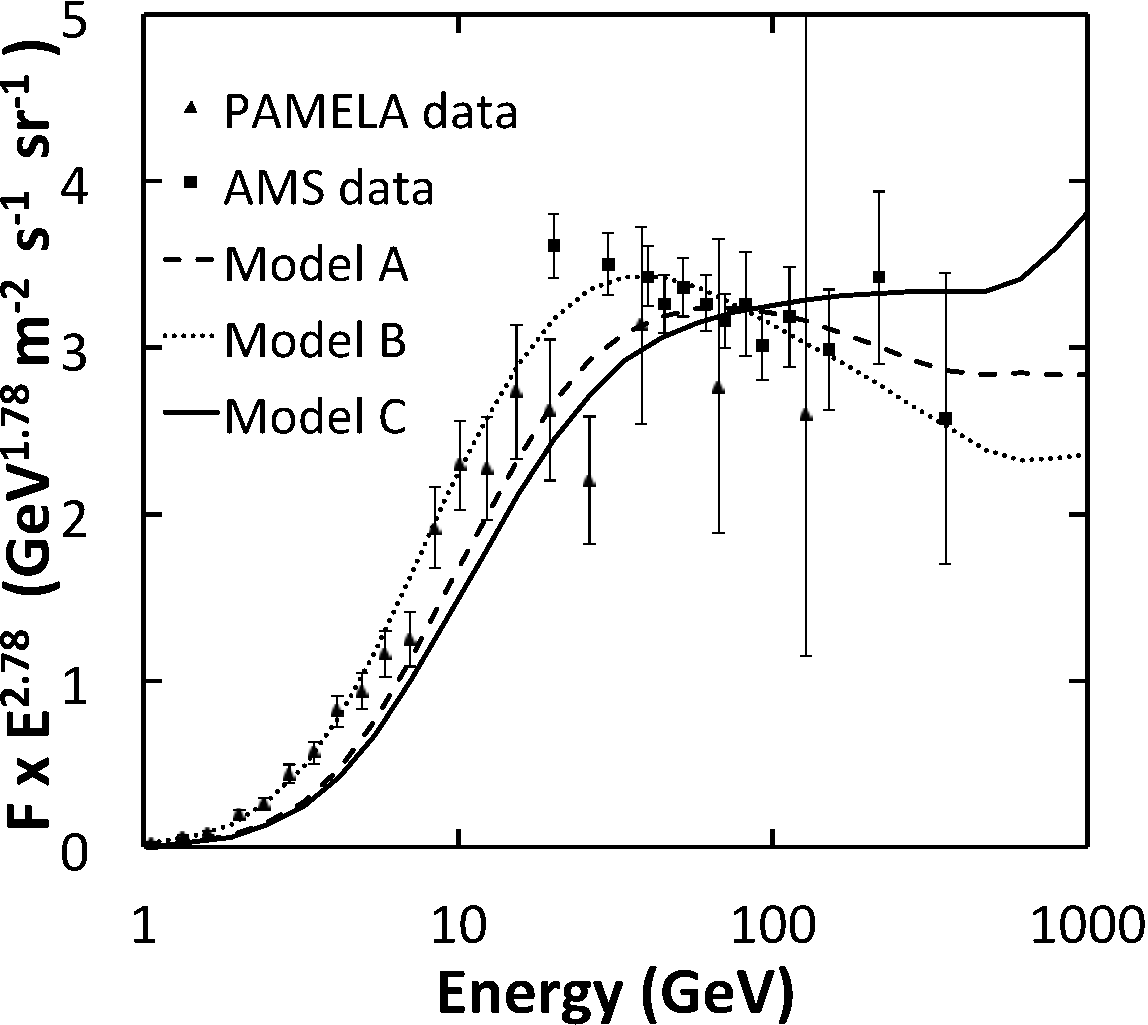}
		\put(55,25){\textbf{\textsf{Anti-Protons}}}
	\end{overpic}
	\begin{overpic}[width=0.38\textwidth]{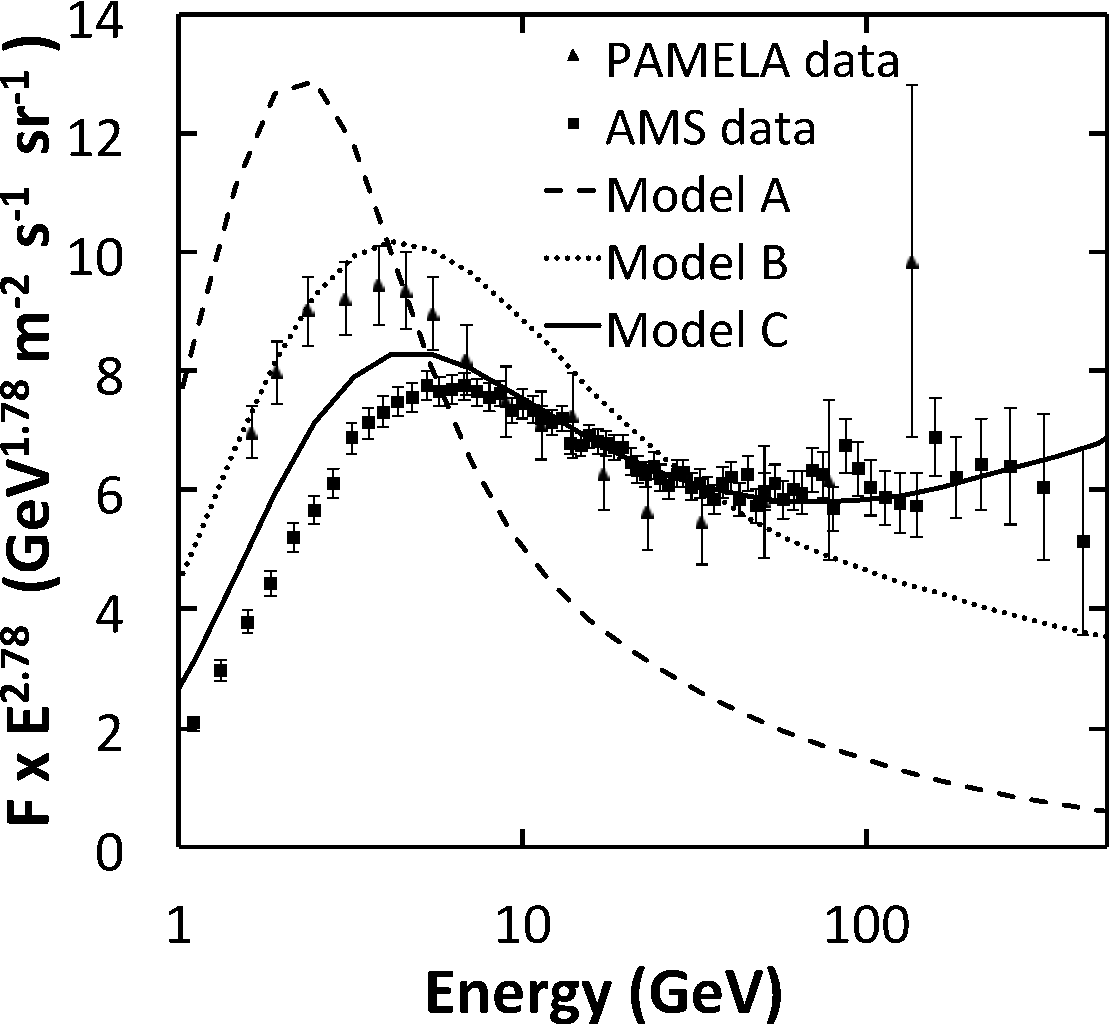}
		\put(25,25){\textbf{\textsf{Positrons}}}
	\end{overpic}
	\\
	\begin{overpic}[width=0.38\textwidth]{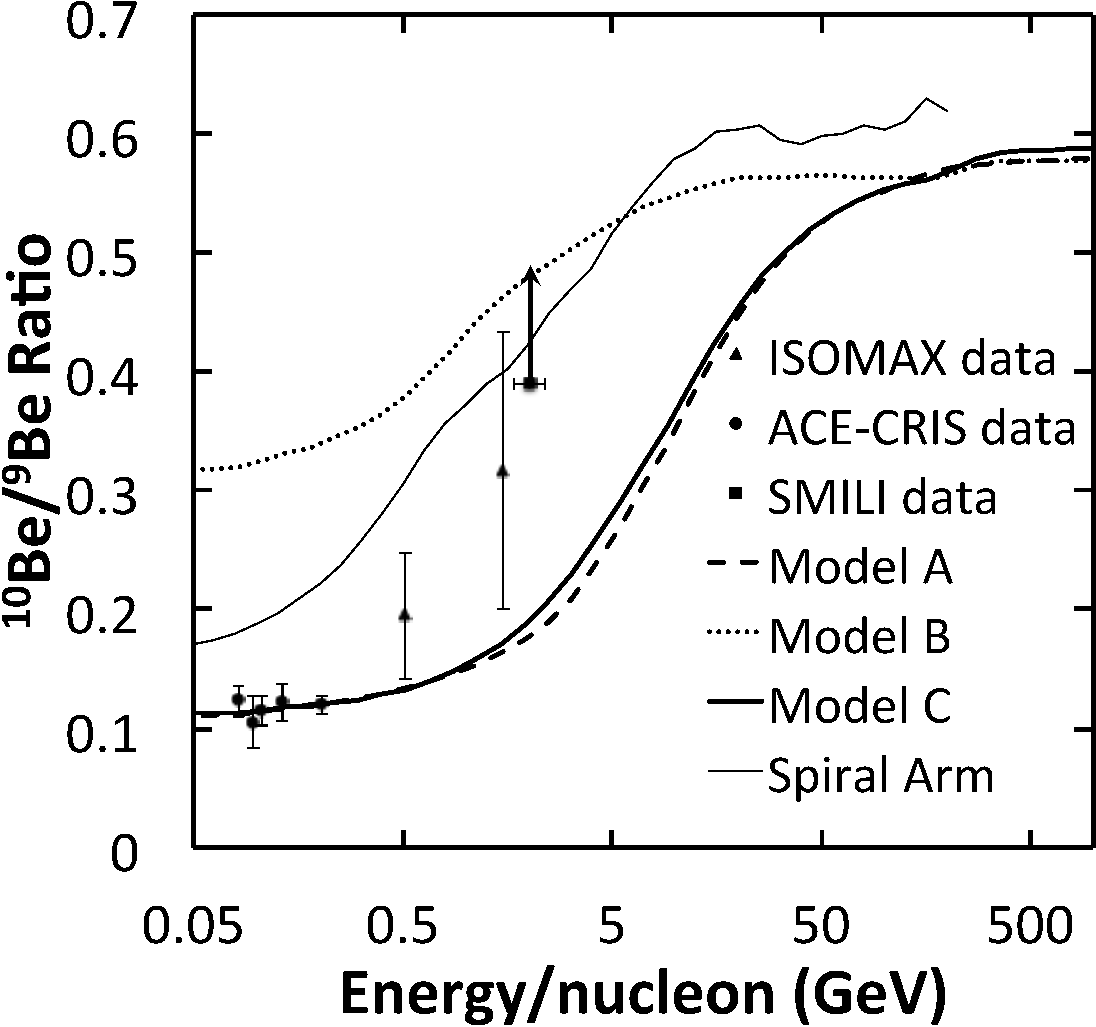}
		\put(30,75){$\dfrac{^{10}\textbf{\textsf{Be}}}{^{9}\textbf{\textsf{Be}}}$}
	\end{overpic}
	\begin{overpic}[width=0.38\textwidth]{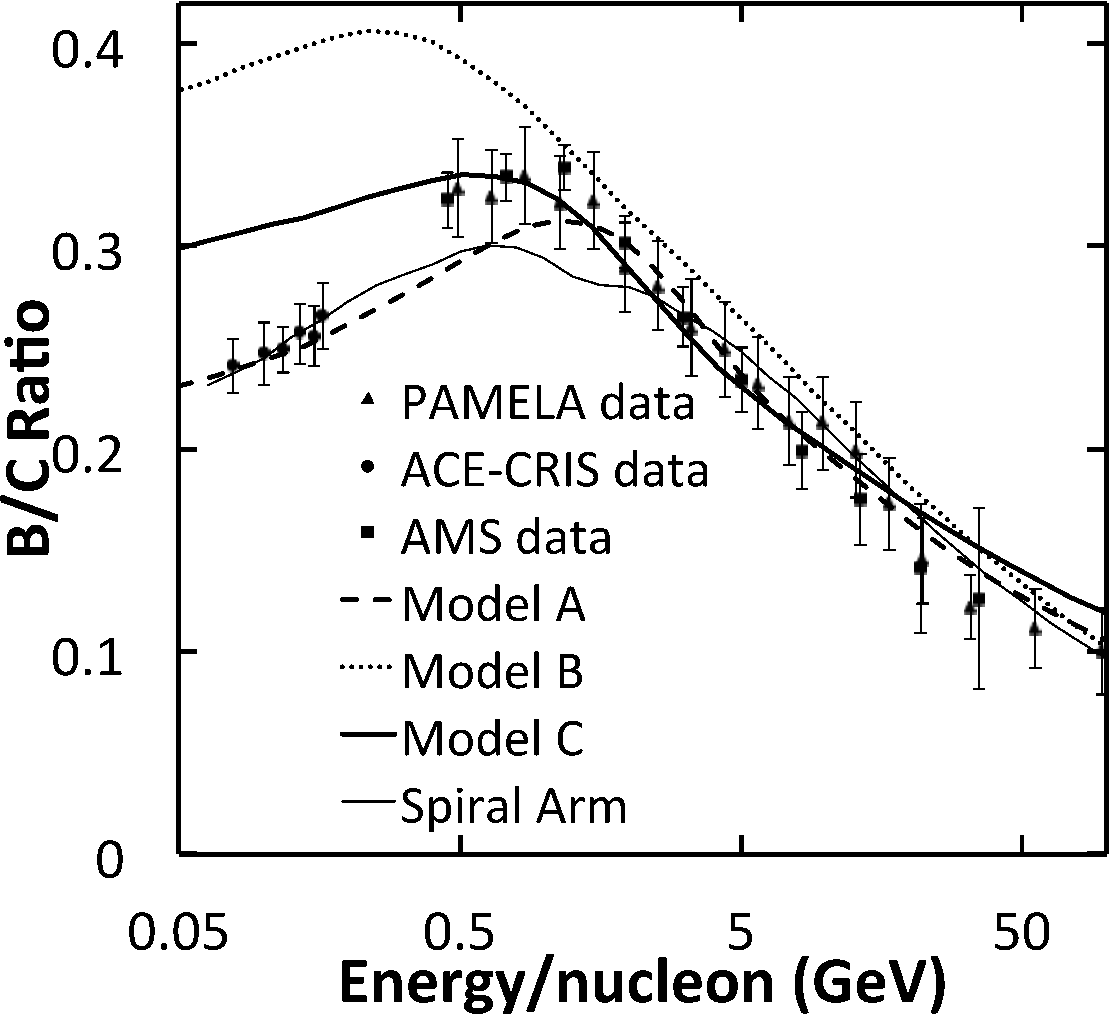}
		\put(65,75){\textbf{\textsf{$\dfrac{\textbf{\textsf{Boron}}}{\textbf{\textsf{Carbon}}}$}}}
	\end{overpic}
	\caption{Comparisons between observed data and the four models considered.  Cosmic ray data were obtained from the public database \citet{database}.  Note that SMILI provides only a lower limit on $^{10}$Be/$^{9}$Be due to the small number of $^{9}$Be and $^{10}$Be events observed (4 and 9 respectively).  The limit shown, which is at the 99\% confidence level, is consistent with the results obtained by ISOMAX.}
	\label{fig:galprop}
\end{figure*}

We obtained promising results with regard to secondary positron production by incorporating Galactic winds, but the results were inconsistent with the B/C ratio at low and high energy. We therefore decided not to pursue such models. Instead we have concentrated on diffusive re-acceleration models, and present here results of calculations of three such models, described below, whose various parameters are summarized in Table \ref{tab:models}, with predictions plotted in Figure \ref{fig:galprop}.

\begin{table*}[!htbp]
  	\centering
	\renewcommand{\arraystretch}{\numexpr\linewidth/\columnwidth}
   	\begin{tabular}{|c|c|c|c|}\hline
      Description & A & B & C \\ \hline
      Nuclear injection index below break 																			& -1.9 & -2.35  & -2.2 \\
      Break in CR injection spectrum (GV) 																			& 11   & - & 11.5 \\
      Nuclear injection index above break 																			& -2.5 & - & -2.55 \\
      Injection offset for heavy (Z $> 1$) nuclei 																& 0.07 & 0.07 & 0.07 \\
      Electron index below first break 																				& 1.6  & 1.6  & 1.6 \\
      First break for electrons (GV) 																						& 2.2  & 2.2  & 2.2 \\
      Electron injection index above first break 																& 2.7  & 3.0 & 3.05 \\
      Second break for electrons (GV) 																				 	& 40   & 40   & 40 \\
      Electron injection index above second break 														  & 2.4  & 2.8 & 2.95 \\
      Alfv\'en speed 																														  & 32   & 0   & 21 \\
      Diffusion coefficient D in $10^{28} \:\text{cm}^2 \:\text{s}^{-1}$ at 4 GV    & 5.75 & 75  & 5.2 \\
      Low energy value of $\delta (D \propto R^\delta)$ 												& 0.3  & 0.52 & 0.28 \\
      Diffusion coefficient break (GV) 																					& 300  & 500  & 500 \\
      High energy value of $\delta (D \propto R^\delta)$ 												& 0.15 & 0.2  & 0.0 \\
      Gas factor																				 												& 1.0  & 30  & 1.0 \\
      Solar Modulation Potential			 (MV)																			& 550  & 550  & 550 \\
      Synchrotron/Inverse Compton (positron) & On & On & Off \\
      Synchrotron/Inverse Compton (electron) & On & On & Off \\ \hline
  \end{tabular}
  \caption{Parameters for GALPROP models}
  \label{tab:models}
\end{table*}

\begin{itemize}
	\item \textbf{Model A} is very similar to the Model P of \cite{Vladimirov}, which is essentially a ``standard'' type model that accounts well for most measurements except for high energy beryllium isotopes and positrons.  A broken power law as a function of cosmic ray rigidity is used to model diffusion, which is $R^{0.15}$ above $300$ GV and $R^{0.3}$ below, with a diffusion coefficient fit to the proton data.  The proton injection spectrum has a spectral index of 1.9 below 11 GV, and 2.5 above.  A nuclear injection offset parameter is included which shifts the spectral index of heavier cosmic ray nuclei by 0.07.  Diffusive reacceleration is included with an Alfv\'en speed of 32 km/s, fit to the measured B/C ratio.  Finally, the electron injection spectrum was fit to match data, with two spectral breaks.
	\item \textbf{Model B} is based off of Model A, but uses a much larger diffusion coefficient to reduce confinement time.  To properly account for the B/C ratio and the observed antiproton flux, the gas density in the ISM is scaled by a floating parameter.  A factor of $30$ is required to reproduce the data, which is interpreted as a very rough estimate of the nested leaky box model of \cite{Cowsik73}.  Here, secondary nuclei are initially produced and confined to very dense regions around supernovae, after which they experience energy independent diffusion in the galaxy.  Fitting the B/C ratio effectively turned diffusive reacceleration off, and thereby raised the diffusion power law to 0.52 below 500 GV and 0.2 above.
	\item \textbf{Model C} takes a different approach, exploring the scenario where confinement time of positrons would be significantly shorter than usually predicted.  This is done by turning off the dominant energy loss mechanisms in the propagation of electrons and positrons: inverse Compton scattering and synchrotron radiation.  This may seem counterintuitive, since this model would \textit{increase} confinement time, but it effectively simulates a situation where most of the observed positrons are formed locally and do not have enough time to radiate significantly.  To test the effect of the positron production cross-section, the default formula was replaced with the charged pion production cross section equations given in \citet{Kamae}.  The differences between the two were found to be negligible in most cases.  Above around 500 GV, the proton data was found to be best fit by energy independent diffusion.  Interestingly, the electron flux could be well fit to data in the absence of radiative processes, with only minor adjustments to the spectral indices.
\end{itemize}

The transport equation solved by GALPROP, from \citep{Strong}, is given by,
\begin{align} 
\dfrac{\partial\psi}{\partial t} &= q + \nabla\cdot\left(D_{xx}\nabla\psi\right) \nonumber \\
&+\dfrac{\partial}{\partial p}\left(p^2D_{pp}\dfrac{\partial}{\partial p}\left(\dfrac{1}{p^2}\psi\right)\right) \nonumber \\
&- \dfrac{\partial}{\partial p}\left(\dfrac{\partial p}{\partial t}\psi\right) - \dfrac{1}{\tau_f}\psi - \dfrac{1}{\tau_r}\psi \label{eq:transport}
\end{align}
where $\psi=\psi_i(\vec{x},p,t)$ is the density per unit of particle momentum at a given point in space and time, for a particle species $i$.  The $q=q_i(\vec{x},p)$ term is a function of space and momentum describing the distribution of sources that create particle $i$ with momentum $p$.  The parameters $\tau_f$ and $\tau_r$ represent the timescale for nuclei fragmentation and radioactive decay, respectively.  The transport equation is solved for all of the primary cosmic rays, which are then used as the source functions for secondaries.  GALPROP is a numerical calculation of this equation, so there is a lot of freedom over its parameters, which can be specified as arbitrary functions of space and momentum.  Some of the parameters relevant to this study are described below.

\begin{itemize}

\item \textbf{Diffusion:} The diffusion coefficient, $D_{xx}$ is generally treated as a function of rigidity with the form,
\begin{equation}
D_{xx} \propto \beta D_0 \left(\dfrac{\rho}{\rho_0}\right)^{\delta}
\end{equation}
where $\rho$ is rigidity, $D_0$ is the diffusion coefficient at some reference rigidity $\rho_0$, and $\delta$ reflects the rigidity dependence of diffusion.  $\rho_0$ is taken to be 400 MV in all models considered.  A break can be included in $D_{xx}$ where $\delta$ changes at some fixed $\rho$.

\item \textbf{Diffusive re-acceleration:} Diffusive re-acceleration is an effect that is believed to occur in which cosmic rays undergo 2nd order Fermi acceleration by scattering off inhomogeneities in the galactic magnetic field.  The coefficient $D_{pp}$ in the transport equation parameterizes this effect, and is defined in terms of $D_{xx}$ and the Alfv\'en velocity $v_A$.  The Alfv\'en velocity represents the speed of Alfv\'en waves propagating through the galactic magnetic field.  This term has a very large effect on secondary/primary ratios, especially at low energies.  Without diffusive re-acceleration an ad-hoc break in the diffusion coefficient is needed to explain the decrease of the B/C ratio at low energies.  However, diffusive re-acceleration models are easily able to explain it by adjusting $v_A$ to match the data.

\item \textbf{Nuclear injection:} This is the spectrum with which nuclei emerge from their source as a function of rigidity, corresponding to the $q$ term in the transport equation.  It has the exponential form $\rho^{-\alpha}$, and can undergo a variable number of breaks at which the spectral index $\alpha$ changes (keeping the spectrum continuous).  At high rigidities, $\gtrsim 100$ GV, the observed spectral index is approximately $\alpha - \delta$.

\item \textbf{Injection offset:} As pointed out in \cite{Vladimirov}, the heavier nuclei appear to have a slightly harder spectrum than protons.  This is included in all of the models by adding an injection offset parameter that is added to the injection index of all nuclei with $Z > 2$.  The value that was found to work best was $0.07$.

\item \textbf{Electron injection:} Electron injection is treated very similarly to nuclear injection by GALPROP.  However, many more spectral breaks are required to match the data (usually 2 or 3).  Since electrons are overwhelmingly from primary production, the electron injection parameters are completely disjoint from everything else.  As Model C shows, the electron spectrum can be made to fit data in virtually any scenario.

\item \textbf{Gas Factor:} This is a parameter added to GALPROP that simply scales all of the gas distributions in the galaxy by a fixed number.  One purpose of this is to approximate a situation in which we are either in a very dense region now, or have been recently.  Another is to approximate the nested leaky box model where most secondary production occurs in the very dense regions of SNRs.

\item \textbf{Radiation:} Electrons and positrons, because they are so light, have more significant radiative losses than nuclei.  At low energies, Coulomb scattering in the ISM and Bremsstrahlung are the dominant radiative losses.  Above about 10 GeV, synchrotron and inverse Compton scattering become dominant.  One way to examine a situation where high energy positrons are secondary particles with a lifetime of a few Myrs is to simply ``turn off" these two sources of energy loss.  Although this has no effect on the nuclei spectra, the predicted positron spectrum should be consistent with short lived secondaries.

\item \textbf{Pion cross section:} An alternative formula was considered for the charged pion production cross section, as calculated in \citet{Kamae}.  Since most secondary positrons originate from pion decays, this could have a significant effect on the positron spectrum.  In general, it was found to be a better fit to data, but not by any drastic amount.

\item \textbf{Solar Modulation:} Although not strictly a parameter of GALPROP, this is passed to the ``plot\underline{{ }}galprop'' Python script packaged with GALPROP to add solar modulation effects to the predicted spectra.  For cosmic rays below $10$ GeV, this can have a huge impact on the observed spectrum.  Because we are mainly interested in higher energies, this was not a large concern.  A modulation potential of $550$ MV was used for all of the calculations, so this should be kept in mind when comparing to different experiments below $10$ GeV.

\end{itemize}

\section{DISCUSSION}

\begin{table*}[!htbp]
  \centering
	\renewcommand{\arraystretch}{\numexpr\linewidth/\columnwidth}
  \footnotesize
   \begin{tabular}{|c|c|c|c|c|c|c|c|c|}\hline
      Model & $p^+$ & $p^-$ & $e^-$ & $e^+$ & Low E B/C & High E B/C & Low E $^{10}$Be/$^9$Be & High E $^{10}$Be/$^9$Be
      \\ \hline
      A   & OK & OK & OK & NO & OK & OK & OK & NO \\
      B   & OK & OK & OK & OK & NO & OK & NO & OK \\
      C   & OK & OK & OK & OK & OK & OK & OK & NO \\
      Arm &  -  &  -  &  -  &  -  &  OK  &  OK  &  OK  & OK \\ \hline
  \end{tabular}
  \caption{Comparison of Model Predictions}
  \label{tab:comparison}
\end{table*}

Table 3 summarizes the differences between the various models. Model A works well except for high energy beryllium and positrons, due to the fact that the confinement time is too large so that $^{10}$Be decays, and the positrons lose energy by synchrotron radiation and inverse Compton scattering. Model C is artificially induced to have short confinement time for positrons by turning off energy loss effects for electrons and positrons. The electrons can be made to agree with the data by adjusting injection parameters. The electron spectral indices above 40 GV for the four models range from 2.4--2.95. These are consistent with the spectral index range of electrons at supernova remnants based on synchrotron observations \citep{Yamazaki}. There are many uncertainties associated with the production and propagation of cosmic ray electrons. For example, \citet{Werner} have considered the effects of including spiral arm source contributions for electrons, and have found the spectrum at Earth is very sensitive to the spiral arm fraction.  They find that the the electron spectrum observed at Earth is strongly dependent on the location of Earth relative to any non-axisymetric distribution of sources.  Indeed, a 10\% spiral arm contribution results in a reduction of 28\% in high energy electrons compared to a model with an axisymmetric source distribution.  They conclude that direct measurements of electron spectra do little to define the primary electron injection spectrum.

The positron spectrum fits remarkably well for Model C with the data. But the $^{10}$Be/$^9$Be ratio does not fit at high energy. Model B fits high energy data reasonably well since propagation parameters were adjusted to make confinement time short for beryllium and positrons. But low energy B/C and $^{10}$Be/$^9$Be do not agree with data.

Since we have been unable to find GALPROP parameters which fit all data, it seems to be the case that GALPROP lacks the ability to model all appropriate physical mechanisms that are crucial for the understanding of cosmic ray secondary production and propagation, or that some measurement/measurements is/are incorrect. \citet{Werner} have noted that two-dimensional azimuthally symmetric modeling of cosmic ray propagation is now giving way to more realistic full three-dimensional modeling as new experimental constraints on the input parameters (matter distributions, magnetic field distributions, cosmic ray source distributions and radiation fields) have become available and computational capability has improved. 

One such new computational procedure that investigates the role of spiral arms in cosmic ray production and propagation is described in \citet{Benyamin}.  They have developed a full three-dimensional simulation that describes cosmic ray diffusion in the Milky Way where significant cosmic ray acceleration originates near dynamic spiral arms.  They note that at low energies the cosmic ray age is dominated by the time since our last passage through the spiral arms and not by diffusion.  They are able to reproduce the observed spectral dependence of secondary to primary ratios  without resorting to additional assumptions (galactic winds, re-acceleration or constraints on diffusivity).  At low energies,  $< 1$ GeV, this results in an increasing secondary to primary ratio as observed.

The results from \citet{Benyamin} for B/C and $^{10}$Be/$^9$Be are shown in Fig.~\ref{fig:galprop} where they are labeled as Spiral Arm. It is interesting to see that they are in reasonable agreement with the large energy independent confinement time at low energy, and short confinement time at high energy, consistent with CRIS, ISOMAX and SMILI-2. Their results for B/C also agree well with the data. This is due to the fact that energy independent confinement time implies increased grammage for higher velocity particles. The mechanism they propose is a diffusion model augmented by a dynamical cosmic ray source caused by the passage of the solar system through a spiral arm over-density region which lasted until about 20 million years ago. Due to the formation of large numbers of short lived hot stars and their subsequent supernovae, this resulted in the injection of cosmic rays into the local ISM. High energy particles are not significantly affected due to their large diffusion coefficient, but low energy particles are effectively trapped, with a relevant time scale of 10--20 million years, consistent with the CRIS lifetime measurement. Modeling of high energy cosmic rays using GALPROP should be accurate, and one can be confident of the results from Model C above several GV.

\section{CONCLUSIONS}

We have shown that the observed rise in the cosmic ray positron fraction can be explained by secondary production without recourse to exotic models involving dark matter annihilation or even sources such as pulsars.  At high energies, positrons, like antiprotons have a spectral index comparable to that of the primary protons at the energies at which they are produced.  This is a natural consequence of secondary production and suggests that antiprotons and positrons are predominantly secondary in origin.  Cosmic ray electrons are predominantly primary in origin and their softer spectrum can be readily understood by a combination of their (unknown) source injection spectrum and energy loss incurred during galactic transport from their sources.  The relatively smaller amount of energy loss for positrons can be understood by the more uniform distribution of their production volume compared to electrons which have a minimum path-length equal to the distance to the closest source.   A decrease of cosmic ray confinement time from 10--15 Myr at 1--2 GeV to 1--2 Myr at 5--6 GeV is required to explain existing measurements of the $^{10}$Be/$^{9}$Be ratio while accounting for the reduction of path length with energy implied by the observed flux of antiprotons.   This feature is not reproduced by simple two-dimensional models of diffusion and re-acceleration such as GALPROP and requires more sophisticated models that incorporate both diffusion and time-dependence in a full three-dimensional framework.  

As cosmic ray and other astrophysical measurements become more precise and are extended to higher energy, it will become possible to create more sophisticated galactic propagation models that incorporate all these features in a truly three-dimensional time-dependent framework.  In particular, accurate Be isotope measurements that extend up to Lorentz factors of 10 will allow the breaking of degeneracy between path-length and confinement time.  More accurate spectral measurements of antiprotons and their proton progenitors will better define breaks in the primary proton spectrum and allow more accurate calculations of the expected positron secondary spectrum.  Once this secondary background of positrons is fully understood, more accurate measurements of the positron spectrum at higher energies may be able to reveal small unexpected and interesting primary components that cannot be resolved by existing measurements. 

\acknowledgments
We acknowledge many useful discussions with Joel Bregman and Dietrich M\"uller and extensive comments by a thorough yet anonymous referee that led to significant improvements in this paper.

\end{document}